\documentclass{article}
\usepackage[T1]{fontenc}
\usepackage{lmodern}
\usepackage{amssymb}

\usepackage{amsthm}
\usepackage{graphicx}

\usepackage{mathnotation}

\usepackage[compress]{natbib}
\bibpunct{[}{]}{;}{a}{,~}{,}

\newcommand{\NOcitep}[2][]{}
\newcommand{\NOcitet}[2][]{}
\usepackage{marginnote}

\newcommand{\concept}[1]{\emph{#1}}

\newcommand{\bvels}[1]{\bodyvel_{\hspace{-1pt}\scriptscriptstyle{#1}}}
\newcommand{\bvelsn}[1]{\hspace{-1pt}\bodyvel_{\scriptscriptstyle{#1}\scriptstyle{,n}}}
\newcommand{\err}[1]{e_{\hspace{-1pt}\scriptscriptstyle{#1}}}

\usepackage{hyperref}
\usepackage{color}

\usepackage{authblk}

\begin{document}

\title{A Data-Driven Approach to Connection Modeling}

\author[1]{Brian Bittner}
\author[2]{Ross L. Hatton}
\author[3]{Shai Revzen}
\affil[1]{Robotics Department, University of Michigan, Ann Arbor, USA \\ babitt@umich.edu}
\affil[2]{Collaborative Robotics and Intelligent Systems (CoRIS) Institute \&
School of Mechanical, Industrial, and Manufacturing Engineering, Oregon State University, Corvallis, USA\\ ross.hatton@oregonstate.edu}
\affil[3]{Electrical Engineering and Computer Science Department  \& Ecology and Evolutionary Biology Department \& Robotics Institute, University of Michigan, Ann Arbor, USA \\ shrevzen@umich.edu}

\maketitle

\newcommand{\Spc}[1]{\mathcal{#1}}
\newcommand{\Xshp}{\Spc{B}}
\newcommand{\Xpos}{\Spc{G}}
\newcommand{\Xalg}{\mathfrak{g}}
\newcommand{\Xcnf}{\Spc{Q}}
\newcommand{\SE}{\mathsf{SE}}
\newcommand{\se}{\mathsf{se}}
\newcommand{\Real}{\mathbb{R}}
\newcommand{\Complex}{\mathbb{C}}
\newcommand{\Qval}{{q}}
\newcommand{\Qx}{{x}}
\newcommand{\Qg}{{g}}
\newcommand{\Qb}{{b}}
\newcommand{\dQval}{{\dot{\Qval}}}
\newcommand{\dQx}{{\dot{\Qx}}}
\newcommand{\dQb}{{\dot{\Qb}}}
\newcommand{\dQg}{{\dot{\Qg}}}
\newcommand{\Sph}[1]{{\mathsf{S}^{#1}}}
\newcommand{\T}{\mathsf{T}}
\newcommand{\D}{\mathsf{D}}
\newcommand{\Img}{\mathrm{Im}}
\newcommand{\A}{\mixedconn}
\newcommand{\dd}{\mathrm{d}}
\newcommand{\ePhi}{\mathrm{e}_\varphi}
\newcommand{\DA}{\mathbf{DA}}

\newcommand{\TBD}{{\color{red}\fbox{TBD}}}

\begin{abstract}
The study of motion in animals and robots has been aided by insights from geometric mechanics.
In friction dominated systems, a mechanical ``connection'' can provide a high fidelity mechanical model.
The connection is a co-vector (Lie algebra) valued map on the configuration space of the system.
As such, empirically estimating a global model of the connection requires a truly exhaustive collection of experiments, and is thus prohibitive on all systems with even a moderate number of degrees of freedom.
In this work, insights from data driven oscillator theory enable us to define a framework for estimating a local model of a connection in the vicinity of observed animal and robot gait cycles.
The estimates are produced directly from motion capture data of a stochastically perturbed cyclic behavior.
We demonstrate the model extraction process under noisy, experiment-like conditions by simulating planar multi-segment serpentine swimmers in a low Reynolds number (viscous-friction) environment.
Following this, we assess model accuracy in the presence of observation error.
Validating our method's capability to produce accurate models in the presence of simulated system and observational noise motivates its usage on real robotic and biological systems.
\normalsize
\end{abstract}

\section{Introduction}
\label{sec::intro}
The ability to move effectively through the environment is both a defining property of animals and a highly desirable capability for man-made systems such as robots and vehicles.
Locomotion (aquatic, terrestrial, and aerial) is most commonly achieved by having a moving body change shape in a way that produces reaction forces from the environment; these reaction forces in turn propel the body.
Robotic and biological systems have similar classes of locomotive goals.
Whether the system wants to achieve some net displacement, or have a general mode for efficient transportation, both systems must exploit their propulsive relationships with the environment.
Commonly, locomotion is achieved via moving the body in such a way that the environmental reaction forces generate a net propulsive motion.

The motivation of this work is to construct high fidelity motion models that account for these propulsive relationships.
Our scope pertains to a class of systems that can be modeled by a ``mechanical connection''.
A subclass of these systems are those whose dynamics are dominated by friction, such that any momentum gained in the system is quickly dissipated.
This paper details a new approach to connection modeling, which leverages data-driven oscillator theory to build a local representation of the connection in the vicinity of observed behaviors.

One of us (Hatton) has developed a framework within the field of geometric mechanics for characterizing gait efficiency in terms of the length and area of the cycle in the shape space \citep{Hatton:2011IJRR, Hatton:2015EPJ, Hatton:2017TRO:Cartography, Ramasamy:2016aa, Ramasamy:2017by}.
These techniques require knowledge of a system's equations of motion; in absence of such an explicit model, this is currently an unsolved problem.
Exhaustive exploration of system dynamics used in ~\citep{Hatton:2013PRL, Dai:2016aa} becomes logistically infeasible when hoping to analyze systems of substantial complexity.
Examining the motion model for animals, whose body motions we cannot control, is also infeasible with this current set of tools.

In the field of data-driven oscillator theory, another of us (Revzen), has developed a set of tools for extracting oscillator-like motion models from noisy and irregularly-spaced data \sloppy{\citep{RevzenPhD, revzen2010fdsd}.}
The methods have an advantageous robustness to the system noise that is intrinsic to real processes, and extends nicely to higher dimensional spaces.
The method does not describe gait adjustments at all, since Floquet models are gait specific.

The geometric and Data-Driven Floquet Analysis (DDFA) approaches can complement each other, addressing some of the weaknesses of both.
By assuming the system has the structure of a mechanical connection, the DDFA tools need only model the specific terms that correspond to connection-like dynamics.
At the same time, DDFA allows the connection modeling approach to drill down the neighborhood of a single gait, exponentially reducing the quantity of data needed for given prediction power.
Restricting attention to local models allows the number of cycles of motion needed to scale linearly with dimension, whereas full connection models would need to scale exponentially with the dimension.

This work details a method for combining the geometric insights of Hatton's work with Revzen's DDFA work.
The DDFA is used to extract a phase-averaged behavior from motion capture data, and compute a local, data-driven connection centered about that behavior.
This leverages intrinsic variability in the system dynamics to understand the dynamical model in the neighborhood of an observed gait cycle.

Below we use simulated mechanical swimming platforms to demonstrate the precision of these data driven geometric mechanics models, and their persistence in high noise environments.
Finally, we point to the utility of the new methods in system identification and field robotics.

\section{Geometry of Locomotion}\label{sec:geolocomotion}

\begin{figure*}[t]
\begin{center}
\includegraphics[width=.9\textwidth]{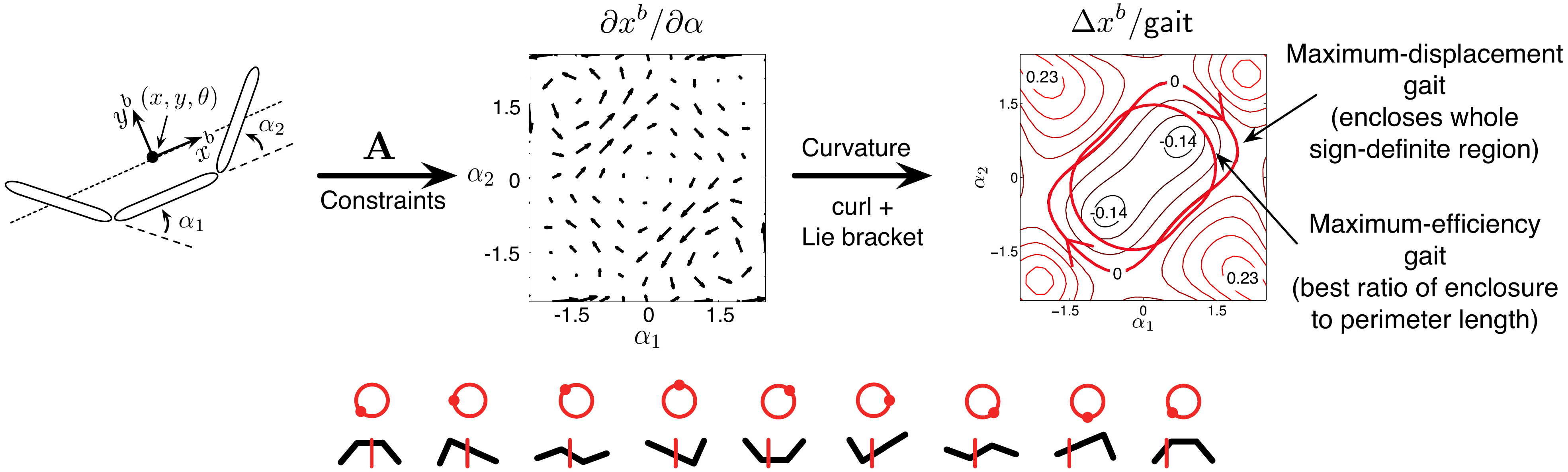}
\caption{%
	Geometric tools used in this work are drawn from \citep{Hatton:2011IJRR, Hatton:2017TRO:Cartography}. %
	System dynamics and constraints placed on the three link swimmer (left)  produce a relationship between shape changes and body position changes. %
	The net displacement that results from a cycle in the shape space (right) corresponds to the amount of curvature in the constraints that the gait encompasses. %
    A time-effort cost is calculated as the path length in the shape space for the cyclic motion.}
\label{fig:geomech}
\end{center}
\end{figure*}

Geometric modeling of locomotion encompasses the first thread of prior work.
Mobile deformable systems can split their configuration space $\bundlespace$ (i.e.\ the space of its generalized coordinates $\bundle$) into a position space $\fiberspace$ and a shape space $\basespace$.
This convenient separation allows for the position $\fiber\in\fiberspace$ to locate the system in the world frame, and the shape $\base\in\basespace$ to give the relative body arrangement that details the localized shape of the platform.\footnote{
$\bundlespace$ is the structure of a trivial, principal \concept{fiber bundle}. $\fiberspace$ is the \concept{fiber space} and $\basespace$ is the \concept{base space}.}

Shape changes provoke reaction forces from the environment, which in turn drive changes in the system's position.
For this work, we use a (geometric) locomotion model
\begin{equation} \label{eq:kinrecon}
\bodyvel
= \mixedconn(\base)\basedot,
\end{equation}
where $\mixedconn$, the \concept{local connection}, is a linear mapping from the shape velocity $\basedot$ to the body velocity $\bodyvel = \fiberinv\fiberdot$ (i.e., the position velocity in the current forward, lateral, and rotational directions of the body frame).
The local connection has a similar function to the Jacobian map for kinematic systems --- it takes the joint velocities to position velocities (here, velocities in the body frame instead of a frame on the end effector).
Both mappings are induced by the constraints applied to the system.

In previous work, the energy cost of changing shape is modeled as a correspondence to the length $\alnth$ of the trajectory through the shape space,
\begin{equation}\label{eq:alnth}
\alnth = \int \sqrt{\transpose{d\base} \metric(\base)\, d\base} = \int_{0}^{T} \sqrt{\transpose{\basedot} \metric(\base)\, \basedot}\ dt,
\end{equation}
where $\metric$ is a Riemannian metric on the shape space that weights the costs of changing shape in various directions.

This connection-and-metric model applies to systems that move by pushing directly against their environment with negligible accumulated momentum in ``gliding'' \sloppy{modes}, and whose energetic costs are governed primarily by internal or external linear (viscous) dissipative effects.
These friction dominated models have been derived analytically for swimmers in low Reynolds number fluids~\citep{Avron:2008, Hatton:2017TRO:Cartography},
and have also been validated experimentally for several robots in dry granular media~\citep{Hatton:2013PRL, Dai:2016aa, McInroe:2016aa}.

The cost encoded by the metric $\metric$ can be thought of as the time it will take the system to complete a maneuver given a unit power budget.
Systems encountering dry friction can compute $\alnth$ as the energy dissipated while executing the motion~\citep{Dai:2016aa}; systems navigating in the viscous friction model we consider in this paper can compute $\alnth$ as the time-integral of the square root of power dissipated~\citep{Hatton:2017TRO:Cartography}.
The connection $\mixedconn$ and metric $\metric$ are important for distinguishing the extremality and efficiency of gaits.

\subsection{Extremal and efficient gaits}\label{sec:gaits}

Mechanical and biological systems typically move via repeated execution of \concept{gaits} -- cycles of shape change -- where each cycle results in a net displacement.
Chained cycles result in additive displacement, producing larger motions in the world.

Geometrically, a gait $\gait$ is a closed shape trajectory with period $T$,
\begin{align}
\gait : [0,T] \rightarrow \basespace & & \gait(0)=\gait(T),
\end{align}
and the system shape at time $t$ during the gait is $\base=\gait(t)$.

The locomotion model in~\eqref{eq:kinrecon} provides that a net displacement over one gait cycle is equal to the path integral of the local connection $\mixedconn$ over that shape trajectory.
This displacement can be approximated\footnote{The approximation quality is dependent on the choice of system's body frame.
This can be optimally selected once $\mixedconn$ is calculated for any chosen frame. For further discussion of this point, see~\citep{Hatton:2011IJRR, Hatton:2013TRO:Swimming, Hatton:2015EPJ}.} as the integral of the \emph{curvature} of $\mixedconn$ over a surface $\gait_{a}$ bounded by the gait,
\begin{align}
\gaitdisp &= \ointctrclockwise_{\gait} \fiber\mixedconn(\base)\, d\base  \approx \iint_{\gait_{a}} \overbrace{\extd \mixedconn +
\textstyle{\sum}\big{[}\mixedconn_{i},\mixedconn_{j>i}\big{]}}^{\text{curvature } D\mixedconn} \label{eq:gaitpathintegral}.
\end{align}
This approximation is made via extension of Stoke's theorem.

The curvature $D\mixedconn$ can formally be expressed as the total Lie bracket or covariant exterior derivative of $\mixedconn$~\citep{Hatton:2015EPJ}.
This curvature is a measure how much the coupling between shape motions and position motions changes across the cycle.
It measures the net displacement that can be extracted from a gait cycle.
Its components $\extd \mixedconn$ and $[\mixedconn_{i},\mixedconn_{j}]$ are the exterior derivative (curl) and local Lie bracket of the system constraints.
The curl captures the net forward-minus-backward motion of a shape motion.
The local lie bracket measures parallel-parking-like motions available to the system -- motions derived from the non-commutativity of constraints.
These measurements are calculated as
\begin{align} \label{eq:extd}
\extd \mixedconn = \sum_{j>i}\left(\frac{\partial \mixedconn_{j}}{\partial \base_{i}} - \frac{\partial \mixedconn_{i}}{\partial \base_{j}}\right)  d\base_{i} \wedge d\base_{j}
\end{align}
and
\begin{align} \label{eq:locallie}
[\mixedconn_{i},\mixedconn_{j}] = \inv{\fiber} \left( \frac{\partial (\fiber\mixedconn_{j})}{\partial \fiber}\mixedconn_{i} - \frac{\partial (\fiber\mixedconn_{i})}{\partial \fiber}\mixedconn_{j} \right) d\base_{i} \wedge d\base_{j}
 = \begin{bmatrix} \mixedconn^{y}_{i}\mixedconn^{\gait}_{j}- \mixedconn^{y}_{j}\mixedconn^{\gait}_{i} \\ \mixedconn^{x}_{j}\mixedconn^{\gait}_{i}- \mixedconn^{x}_{i}\mixedconn^{\gait}_{j} \\ 0 \end{bmatrix} d\base_{i} \wedge d\base_{j},
\end{align}
where the wedge product $d\base_{i} \wedge d\base_{j}$ is the basis area spanned by basis vectors indexed $i$ and $j$.

If a system has a two degree of freedom shape space, $\extd \mixedconn$ and $[\mixedconn_{i},\mixedconn_{j}]$ have only a single component (on the $d\base_{1} \wedge d\base_{2}$ plane), and~\eqref{eq:gaitpathintegral} reduces to an area integral.
This integrand provides the magnitude of $D\mixedconn$.
A candidate goal function for systems is to maximize net displacement per cycle.
These extremal gaits lie along zero-contours of $D\mixedconn$, maximizing the area of the sign-definite region they enclose on the vector field.

Generally, extremal gaits have more value as mathematical objects than as desirable behaviors in robots and animals.
With the exception of sports such as basketball, where step counting provides special incentives for long steps, displacement-per-cycle is not a useful goal function for locomotion.
It leads to wasted time or energy with respect to other motions that could achieve the same displacement in a larger number of cycles.
\footnote{%
	As discussed in~\citep{Tam:2008}, defining a ``cycle'' in this context also introduces its own ambiguities --- does a motion that almost returns to its starting shape, then makes another loop count as one cycle or two? This problem is especially acute for systems with more than two shape variables, with the higher dimensional shape spaces admitting gait curves such as helices that have many almost-identical sub-cycles, but never cross themselves.}
A better locomotive goal function is efficiency, which is calculated by dividing the displacement per cycle by the effort or time required to execute it, producing a measure of gait efficiency.

In our model, efficiency $\gamma$ is the ratio between the net displacement $\gaitdisp$ it induces and the path-length cost $\alnth$ calculated in~\eqref{eq:alnth}, $\gamma := \frac{\gaitdisp}{\alnth}$.
Maximizing this efficiency can be thought of as maximizing speed at a given power (or minimizing power at a given speed).
Efficient gaits are thus always the most desirable for effective locomotion in power limited systems, regardless of whether the desired characteristic is to ``move fast'' or ``move efficiently.''

\subsection{Empirical geometric models}\label{sec:empiricalgeometric}

This geometric approach was intended to be used for analysis of systems that could be modeled from first principles to have the form in ~\eqref{eq:kinrecon}.
In ~\citep{Hatton:2013PRL, Dai:2016aa} we built on this work to demonstrate that the constraint curvature $D\mixedconn$ can be used for inspection of motion models for systems where the dynamics are less ``clean,'' and are only computable through numerical modeling and simulation or empirical observation.

Nonlinear models~\citep{Hatton:2013PRL} or experimental measurements~\citep{Dai:2016aa} were first used to sample the relationship between $\fibercirc$ and $\basedot$ across the tangent bundle $T\basespace$.
A linear fit was assigned to this relationship on a grid of tangent space base-points $T_{\base}\basespace$, giving $\mixedconn$ on a sampling of the shape space.
From this we then calculated the components of $D\mixedconn$ as per~\eqref{eq:extd} and~\eqref{eq:locallie}.
We were then able to directly identify effective gaits for translation and rotation (of a three-link and serpenoid system) by plotting the curvature over the shape space.
This follows the illustrated process in Fig.~\ref{fig:geomech}.

\begin{figure*}
\begin{center}
\def\svgwidth{.96\textwidth}
\input{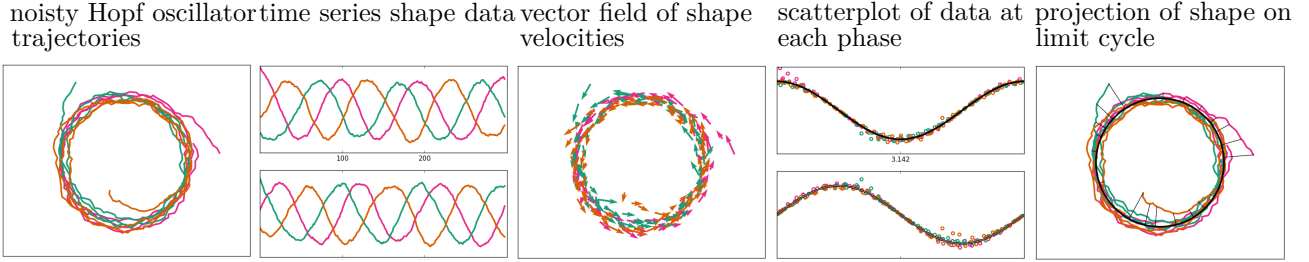}\label{fig:hopf}
\caption{%
	Phase estimation tools are applied to a Hopf oscillator via Data-Driven Floquet Analysis. %
    The shape data are seen to converge to a noisy circle (extreme left) and are shown as a periodic time series (second from left). %
    The differentiated shape signals produce a vector field of shape velocities (middle). %
    The limit cycle to the data is computed using a phase estimator \citet{revzen2010fdsd}. %
    This estimation provides a cycle as a function of phase (black lines on right plots), along with a canonical map from shape data to asymptotic phase on a limit cycle. %
    Surfaces of constant phase (isochrons) arise as radial lines in Hopf oscillators.
}
\end{center}
\end{figure*}

\section{Oscillators and Data-driven Modeling}\label{sec:oscillators}

A robust theory of gaits as oscillators, combined with a statistical approach to data-driven model construction makes up the second thread of prior work.
For observed systems, it is not always known beforehand what the limit cycle is, what the dynamics of attraction to the limit cycle are, or even how long the period of oscillation is.
Data-Driven Floquet Analysis (DDFA)~\citep{RevzenPhD, revzen2015_SPIE} enables extraction of these properties.
The key elements of this extraction process are specified below.

To provide a simple example, we will assume all observation come in a single regularly sampled time series.
These observations consist of numerically differentiable  $(\fiber_{n}, \base_{n})$ position and shape samples.
Differentiation might be achieved via a second order Kalman smoother \citep{rauch1965mle, roweis1999unifying}) to pair the samples with velocities $\fiberdot_{n}$, $\basedot_{n}$, and $\fibercirc_{n}=\fiberinv_{n}\fiberdot_{n}$.
From oscillator theory~\citep{oscillationsGuckenheimer, revzen2015_SPIE} we know that any exponentially stable oscillator (which we assume this to be) can be parameterized by phase $\gaitphase:\,\basespace \to [0,T)\subset\Real$ based on the following rules:

\begin{enumerate}
\item Each point on the limit cycle has a unique phase value, and these are spaced such that trajectories on the limit cycle grow in phase at rate $\dot{\gaitphase}=1$.
\item Any point off of the limit cycle has a unique point of equivalent asymptotic phase on the limit cycle.
These two trajectories will coalesce on the limit cycle in finite time.
An \concept{isochron} is the set of all points sharing the same asymptotic phase of the oscillator.
The oscillator trajectories advance across these isochrons such that $\dot{\gaitphase}=1$ at all points on the isochron.
\end{enumerate}
Our modeling process was as follows: each sample $n$ was assigned a phase $\gaitphase_{n}$ via a \concept{phase estimator} such as \citet{RevGuk08}, which takes multivariate time-series oscillator data and estimates a phase for each point.
A visual example of the process is provided in Fig. \ref{fig:hopf}.
Once each is sample is assigned a phase, the limit cycle (nominal-gait-as-executed) is modeled by computing a pair of Fourier series $\gait_{0}$ and $\omega$ with respect to the phase: $\gait_0(\gaitphase_n) \approx \base_{n}$ was fitted to the shape data, and $\omega(\gaitphase_n) \approx \basedot_{n}$ was fitted to the shape velocity data.
Elements $\gait_0$ and $\omega$ are computed from separate noisy datasets, so the condition $\dot{\gait_0}=\omega$ need not be satisfied after this fitting procedure.
We create a self-consistent model $\gait$ of the limit cycle by using a matched filter to combine the analytical integral of $\omega$ with the $\gait_0$ estimate.
This allows us to obtain a single self-consistent cyclic trajectory.
Past experience \citep{RevzenPhD} has shown that this self consistent model is a more accurate representation of the limit cycle than the shape model $\gait_0$ directly fitted from noisy data.

\section{Data-Driven Modeling of the Connection }\label{sec:method}

The gait analysis methods described in \S\ref{sec:geolocomotion} provide a powerful link between gaits' geometry and performance characteristics such as extremality and efficiency.
Their usage requires having a model for how small shape changes induce body motion changes.
For systems that experience nontrivial, complex interactions with their environment, there exists no closed form procedure to extract their models from first principles (even if their net effect can be modeled as the linear relationship in~\eqref{eq:kinrecon}), and exhaustive empirical evaluations~\citep{Dai:2016aa} become infeasible as we move to system that has many shape variables and/or cannot execute desired motions due to limited control capabilities.

Conversely, DDFA as described in~\S\ref{sec:oscillators} can compute meaningful models from noisy measurements by characterizing the system as an oscillator.
However, the model is both local (so specific to the gait) and does not provide context for comparing the gait against new candidate motions.

The main contribution of this current work is based on the observation that the data-driven modeling approach can allow us to quickly compute an approximation of the connection in a tube around any observed cycle, or behavior.
This first order approximation of the connection allows us to rapidly compute the dynamics of \textit{any} gaits that lie within this tube.

In this innovation, we exploit that $D\mixedconn$ is a two-form and thus a linear map.
It can be reconstructed at every point along a gait cycle using a set of phase-centered regressions applied to the relationship between $\fiber$ and $\base$ collected via observation of behaviors.

\subsection{Analytic Approximation of the Connection Near a Gait}
\label{sec:gaitCentered}

In this section, we present an approximation of the mechanical connection and the cost metric.
Each is centered about a nominal gait.
We then detail a procedure for estimation of the local model elements from data.

As discussed in \S\ref{sec:oscillators}, DDFA can fit a gait cycle $\gait(\cdot)$ to observed shape data $r$.
Shape data off of this limit cycle are written as perturbations $\delta(t) := \base(t) - \gait(t)$.
Using this notation for perturbations, we can write $\mixedconn(\cdot)$ in a neighborhood of the point-set $\text{Im}\,\gait$ using its Taylor series,
\begin{align}
\label{eqn:diffmdl}
\mixedconn^{k}_{i}(\base)\dot{\base}^{i} &=
\mixedconn^{k}_{i}(\gait+\delta)\dot{\base}^{i} \approx \left[
	\mixedconn^{k}_{i}(\gait)
	+\frac{\partial\mixedconn^{k}_{i}}{\partial\base_{j}}(\gait)\delta^{j}
\right]\dot{\base}^{i},
\end{align} where, as per Einstein index notation, $\mixedconn_{i}^{k}$ corresponds to the element in the $k$-th row and $i$-th column of $\mixedconn$.
Writing the derivative of the connection across the shape space allows us to estimate the connection for shape data that doesn't lie explicitly on the cycle.

Note that $\frac{\partial\mixedconn^k}{\partial \base}$ is not the Hessian matrix of $\fiber^k$ with respect to $\base$ around points on the gait, i.e. it is not a double gradient.
Computation of the Hessian would require $\fiber$ to be a function of $\base$, but it is not.
If such a function existed, gait cycles would return zero net displacement. The asymmetry of $\extd\mixedconn^k$ in~\eqref{eq:extd} directly measures the system's ability to locomote along the $k$-th direction.
Similarly, the $[\mixedconn_{i},\mixedconn_{j}]$ term from~\eqref{eq:locallie} measures the covariant asymmetry of $\frac{\partial \fiber \mixedconn}{\partial \fiber}$. This covers expansion of the connection from local to global coordinates.

\subsection{Estimating $\mixedconn(\gait)$ and $\DA(\gait)$ from data}
\label{sec:estimatingA}

\noindent
We begin our system identification process by applying the gait extraction algorithm described in \S\ref{sec:oscillators} to input data as a time series of the system shape $\base_n$, shape velocity $\basedot_n$, and observed body velocity $\fibercirc_{n}$, where we have N time points.
Fourier series models of $\gait(\cdot)$ and of $\gaitv(\cdot)$ are produced from the data.
We then partition phase into $M$ evenly spaced values, $\gaitphase_{_1} \ldots \gaitphase_{_M}$, to obtain $\gait_{m} := \gait(\gaitphase_m)$ and $\gaitv_m := \gaitv(\gaitphase_m)$.
These shapes and shape velocities fall precisely on the limit cycle.

These become the center points for a first order approximation of the connection.

At each $\gait_m$, all nearby shapes $\base_n$, i.e. $n$ such that $\|\base_n - \gait_m\|<\delta_{\text{max}}$ are collected.
For notational simplicity, when both index $n$ and index $m$ appear the below equations, the values of $n$ are to be restricted such that they lie sufficiently close to the proper center point $\gait_m$ on the limit cycle.
The difference between a shape sample and an on-gait reference point is defined as $\delta_n := \base_n - \gait_m$.

Within each neighborhood of $\gait_{m}$, we use a linear regression to determine the slopes of the relationship between $\bodyvel$, $\basedot$, and $\delta$.
This allows us to estimate the local connection and its derivatives.
This regression is the solution to the Generalized Linear Model that forms by expanding a first order Taylor-series $\mixedconn$ from~\eqref{eqn:diffmdl} into the locomotion model from~\eqref{eq:kinrecon}:
\begin{align} \label{eq:datadriven}
\bodyvel^k_{n} &\sim \left( \mixedconn^k_i \right) \basedot^i_{n} + \left(\frac{\partial\mixedconn^k_i}{\partial\base_j}\right) \delta^j_n \basedot^i_{n},
\end{align}
where $\left( \mixedconn^k_i \right)$ are the $M$ separate estimates of $\mixedconn^k_i(\gait_m)$ and $\left(\frac{\partial\mixedconn^k_i}{\partial\base_j}\right)$ are the $M$ separate estimates of $\frac{\partial\mixedconn^k_i}{\partial\base_j}(\gait_m)$.

When applied to oscillator systems such as that illustrated in Fig.~\ref{fig:hopf}, this straightforward regression is biased by the shape velocity samples being centered around $\basedot = \gaitv_{m}$ rather than $\basedot = 0$.
By re-centering the regression around $\mixedconn(\gait_{m})\gaitv_{m}$, this bias is corrected.
By defining ${\dot \delta}_n := \basedot_n - \gaitv_m$, we can expand the regression to separate the influence of shape velocities off of the Cycle from velocities that adhere to the limit cycle.
To do this we expand the GLM of~\eqref{eq:datadriven} as (for velocity component $k$ and each value of $m$):
\begin{align} \label{eq:advanceddatadriven}
\bodyvel^k_{n} \sim
	& \mcConst^k +  \mcOfs^k_j \delta^j_n
   + \left( \mixedconn^k_i \right) {\dot \delta}^i_{n} + \left(\frac{\partial\mixedconn^k_i}{\partial\base_j}\right) \delta^j_n {\dot \delta}^i_n
\end{align}
where $\mcConst^k := \mixedconn_i^k \gaitv^i$ is the connection applied to the (unmodified) gait cycle shape velocity, and $\mcOfs^k_j := \frac{\partial\mixedconn^k_i}{\partial\base_j}\gaitv^i$ is the combined interaction of shape offset and shape velocity applied to the (unmodified) gait cycle shape velocity.
Here $\mcConst^k$ is a constant (with $k$, $m$ fixed); and $\mcOfs^k$ is a (``co-'')vector that acts on shape perturbations.
The $\left( \mixedconn^k_i \right)$ element is a true co-vector that acts on velocity offsets away from the typical gait velocity.
Finally, an interaction matrix $\left(\frac{\partial\mixedconn^k_i}{\partial\base_j}\right)$  corresponds to shape offsets and shape velocity offsets.

We pose the regression as a least-squares problem (for each $k$ and $m$; indices $k$ and $m$ elided below for clarity):
\begin{equation}\label{eq:advancedregression}
	\begin{bmatrix} \fibercirc_{_1} \\ \vdots \\ \fibercirc_{_N} \end{bmatrix} =
    \begin{bmatrix}
	1,  &  \delta_{_1}, & {\dot\delta}_{_1}, & {\dot\delta}_{_1}\otimes\delta_{_1} \\
	\vdots & \vdots & \vdots & \vdots  \\
    1,  &  \delta_{_N}, & {\dot\delta}_{_N}, & {\dot\delta}_{_N}\otimes\delta_{_N} \end{bmatrix}
\cdot
    \transpose{\begin{bmatrix}
	\widehat{\mcConst},  &
    \widehat{\mcOfs_j}, &
    \widehat{\mixedconn_i}, &
    \widehat{\frac{\partial\mixedconn_i}{\partial\base_j}}
	\end{bmatrix}}
\end{equation}
where $\widehat{~}$ indicates ``estimated'' and $\otimes$ is the outer product.
For a $d$ dimensional shape space, the unknowns on the right have $1+d+d+d^2$ elements.

A model is obtained at each $m$, at which point a Fourier series is fit to the GLM as a function of phase.
This allows for smooth interpolation of the model in phase.

\subsection{Estimating the Metric} \label{sec:estimatingM}

The Riemannian effort-metric $\metric$ can be estimated in a similar manner to the estimation of $\mixedconn$.
This is estimated by recording the differential cost of motion $\dot{\alnth}$ along with the system shape and shape velocities, and then fitting these costs to a linearized expansion of~\eqref{eq:alnth} taken at the phase-partitioned points $\gait_m$ using the matching $n$ indices,
\begin{align} \label{eq:metricregression}
\dot{\alnth}^2_n \sim \transpose{\basedot_n} \left[ \metric + \left(\frac{\partial\metric}{\partial\base_j}\right)\delta^j_n \right] \basedot_n.
\end{align}
This regression is also biased because ofthe $\basedot$ values being centered around $\gaitv$ instead of $0;$. The regression for $\metric$ is re-centered here just as in~\eqref{eq:advancedregression}.
\begin{align} \label{eq:advancedmetricregression}
\dot{\alnth}^2_n \sim (\transpose{\gaitv_n + {\dot\delta}_n}) \left[ \metric + \left(\frac{\partial\metric}{\partial\base_j}\right)\delta^j_n \right] (\gaitv_n + {\dot\delta}_n)
\end{align}
leading to the regression:
\begin{equation}
	\begin{bmatrix} \dot{\alnth}^2_{_1} \\ \vdots \\ \dot{\alnth}^2_{_N} \end{bmatrix} =
    \begin{bmatrix}
	    1, & {\dot\delta}_1,
        & {\dot\delta}_{_1} \hat{\otimes}\, {\dot\delta}_{_1}, & \delta_1, &
        {\delta}_{_1} \otimes\, {\dot\delta}_{_1}, & {\delta}_{_1} \otimes\, {\dot\delta}_{_1} \hat{\otimes} {\dot\delta}_{_1}\\
	\vdots & \vdots & \vdots & \vdots & \vdots \\
	    1,  & {\dot\delta}_N, & {\dot\delta}_{_N} \hat{\otimes}\, {\dot\delta}_{_N},  & \delta_N,
        & {\delta}_{_N} \otimes\, {\dot\delta}_{_N}, & {\delta}_{_N} \otimes\,  {\dot\delta}_{_N} \hat{\otimes} {\dot\delta}_{_N} \end{bmatrix}
\cdot
    \transpose{R},
\end{equation}
\begin{equation}
R =
\begin{bmatrix}
	\widehat{\metric_{i,j}\gaitv^i\gaitv^j}, &
    \widehat{\metric_{i,j} \gaitv^i},& \widehat{\metric_{i,j}}, & \widehat{\frac{\partial\metric_{i,j}}{\partial\base_k} \gaitv^i\gaitv^j}, & \widehat{\frac{\partial\metric_{i,j}}{\partial\base_k} \gaitv^j}, & \widehat{\frac{\partial\metric_{i,j}}{\partial\base_k}}
	\end{bmatrix}
\end{equation}

Additionally, because $\metric$ is a symmetric tensor, only $\binom{d}{2}$ elements are estimated.
This eliminates about half of the regressors and avoids redundancy issues.

\subsection{Comparison of Estimates to Previous Work}

This process is similar to in~\citep{Hatton:2013PRL, Dai:2016aa}, where an empirical estimate of $\mixedconn$ is presented.
This approach offers some distinct advantages.

The previous method required shape velocity samples to identify $\mixedconn$ at a point to lie in the tangent space at that point.
This requirement is relaxed in the presented approach by fitting to a linearized expansion of~\eqref{eq:kinrecon} instead of strictly~\eqref{eq:kinrecon} itself.

Furthermore, the behavior is leveraging the intrinsic noise in the system to build the local model.
From an empirical standpoint, it is convenient to observe a behavior and automatically extract a model fit to a tube shaped neighborhood about the average observed behavior.
In the results presented, there will be cases with system noise and perfect observation, followed by system noise and imperfect observation.

\section{Performance of the Data-Driven Models}\label{sec:accuracy}

Here we benchmark the accuracy of the local models generated by the data-driven geometric approach.
We compared the predicted body velocity for a test system against three system models that had various levels of knowledge about the ``true'' simulated system dynamics.
The test system had a geometric locomotion model of the form in~\eqref{eq:kinrecon}, and its shape trajectories were perturbed by a stochastic process like that illustrated in Fig.~\ref{fig:hopf}.

\subsection{Reference models} \label{sec:referencemodels}

We used the data-driven process described in~\S\ref{sec:method} to construct a phase varying first order model of $\mixedconn$ at points $\gait_m$ along our observed limit cycle.
Each $\base_{n}$ data point from the (noisy) trial was assigned a phase-matched value $\gait_{n}$ on the limit cycle,\footnote{These phase-matched $\gait_{n}$ points can be individually computed for each $\base_{n}$ These values are not constrained to be one of the phase partitioned values $\gait_{m}$.
Similarly, the estimates of $\mixedconn$ and its derivative from \S\ref{sec:gaitCentered} are computed as Fourier series, so can be sampled as smooth functions of phase.
} which allowed us to compare a variety of body velocity models:

\begin{enumerate}
\item The \concept{ground truth model}
\begin{equation} \label{eq:groundtruthbodyvel}
\bvelsn{G} = \mixedconn(\base_{n})\basedot_{n},
\end{equation}
is computed by passing points $(\base_{n},\basedot_{n})$ to the simulator.
\item The fully \concept{data driven model} that is presented above.
Regression estimates of the Taylor expansion of $\mixedconn$ are used to approximate $\mixedconn$ at points nearby the gait cycle, and $\bvelsn{D}$ is given by \eqref{eq:advanceddatadriven}, used with the quantities estimated from \eqref{eq:advancedregression}.
\item An \concept{analytic model}
\begin{equation}  \label{eq:analyticalbodyvel}
\bvelsn{A} = \mixedconn(\gait_n)\dot{\base}_n + \frac{\partial\mixedconn}{\partial\base}(\gait_n)\delta_n\dot{\base}_n
\end{equation}
that uses a first order Taylor-series expansion of the simulator dynamics computed at the same point as the data driven model.
No regression data is used for this computation.
This model gives us a basis to check if the regression values in the data driven model are relatively accurate.
\item A \concept{template projection model}
\begin{equation}  \label{eq:templatebodyvel}
\bvelsn{T} = \mixedconn(\gait_n)\dot{\gait}_n.
\end{equation}
that projects each $(\base_{n},\basedot_{n})$ data point onto the limit cycle that was used to center the data driven model.
This approximation tests how much additional information is gained when using the first order term of the Taylor expansion.
\end{enumerate}

One comparison is that the leading term of the analytical approximation is the template approximation in~\eqref{eq:templatebodyvel} (after separating $\basedot_{n}$ into $\dot{\gait}_{n}$ and $\dot{\delta}_{n}$ components), and that the insights required to predict behaviors induced by shifting the limit cycle are contained in the partial derivatives \eqref{eq:advanceddatadriven} and~\eqref{eq:analyticalbodyvel}.

\subsection{Simulation Setup: Swimming with System Noise}

A three-link Purcell swimmer~\citep{Purcell:1977} modeled as described in \citet{hatton2013geometric} is the baseline platform used for our modeling comparison.
This system is friction dominated, moveing through a viscous fluid with linear drag.
A $2:1$ lateral/longitudinal ratio defines the ratio of forces onto and along the joint connected members of the swimmer.
To demonstrate our methods natural extension to higher dimensions, we also analyzed model accuracy for a nine-link swimmer.
Both are pictured in Fig. \ref{fig:gait} part A2 and A8 (2 degrees of freedom and 8 degress of freedom respectively).

To simulate system noise, shape trajectories are generated as sample paths of a (Stratonovich) stochastic differential equation, injected into the shape space:
\begin{align}\label{eq:simsde}
d \gaitphase &= 1\, d t + \eta \circ d W_\gait, &
d \delta &= -(\alpha\,\delta)\, d t + \eta \circ d W_\delta,  &
\base(t) &:= \gait_{_\text{REF}}(\gaitphase(t)) + \delta(t).
\end{align}
where $\gait_{_\text{REF}}(\cdot)$ was a reference motion on the limit cycle that can be smoothly interpolated in phase; $\alpha$ was the coefficient of attraction pulling off cycle behavior back to the limit cycle; and $\eta$ was a noise magnifier for the Weiner processes $d W$. This process drives noise in both shape and phase.

For all simulations in this paper $\alpha = 0.05$ and $\eta = 0.025$, chosen based on relative geometric similarity to noisy data we have encountered in robotic and biological experiments.

\subsection{Model Accuracy Results}

\begin{figure*}
\begin{center}
\def\svgwidth{\textwidth}
\input{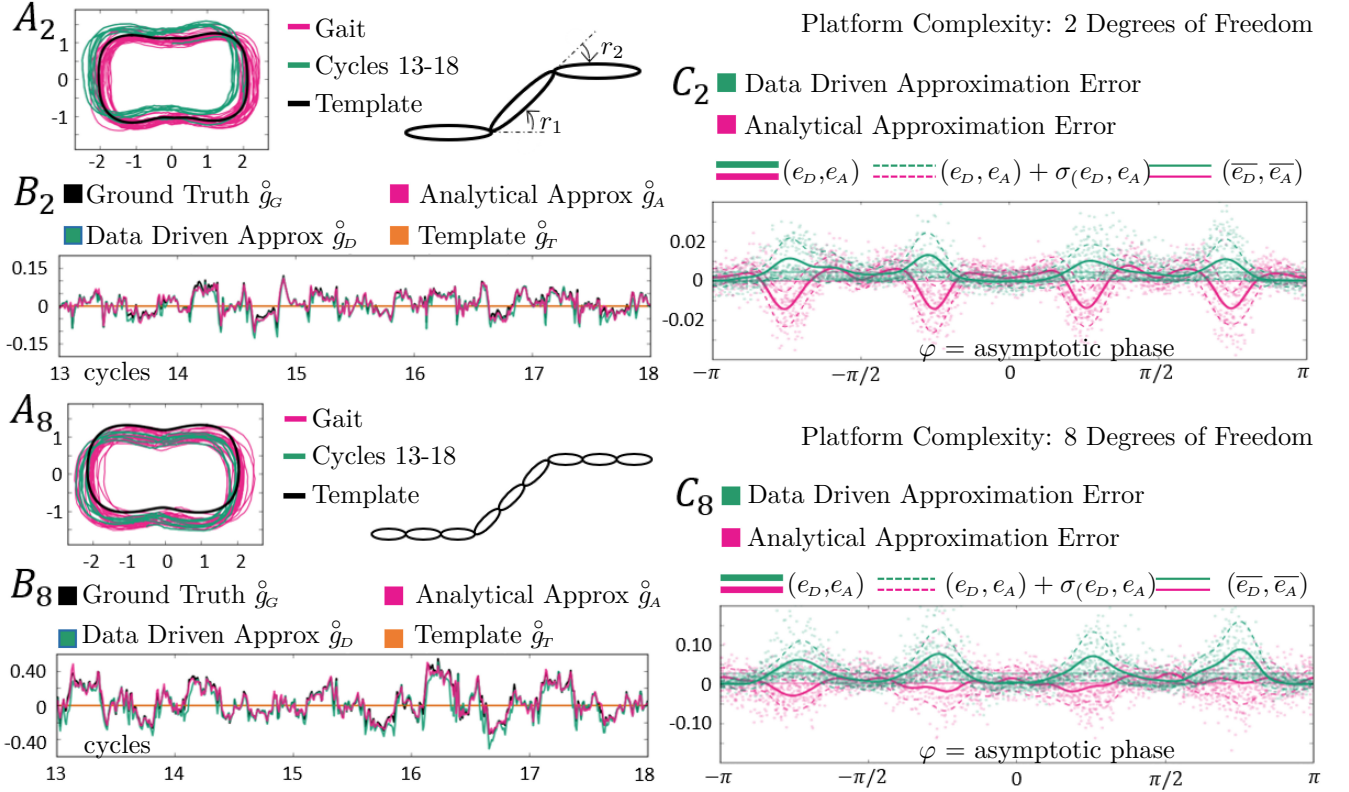}\label{fig:gait}
\caption{ The 3 link and 9 link swimmers are compared for model accuracy via prediction of body velocity.%
  [A] Each platform is perturbed while following an extremal gait (green and magenta; plotted on the first two principal components). The system undulates for 30 cycles. %
  Cycles 13-18 (green) are plotted with corresponding body velocity in part [B], %
  showing the ground truth model (black), the data driven model (green), and the analytic model (magenta) with the template dynamics subtracted (orange). %
  [C] shows the body velocity error of the data-driven and analytical approximations as a function of phase. A feature exclusive to the data-driven model is having zero mean error. %
  The mean error at each phase (solid) becomes lower in the data-driven approximation with respect to the analytical approximation as the number of joints in the swimmer increases.
  }
\end{center}
\end{figure*}

We examine the performance of the data driven model on the extremal gait maximizing motion in the $x$ direction, known from~\citet{Tam:2007, Hatton:2013TRO:Swimming}.
We chose an extremal behavior, since this should be most difficult to capture. Non-extremal motions have strong first order effects in their neighborhood.
Results are presented in Fig.~\ref{fig:gait}.

\begin{figure}
\def\svgwidth{17.6cm}
\input{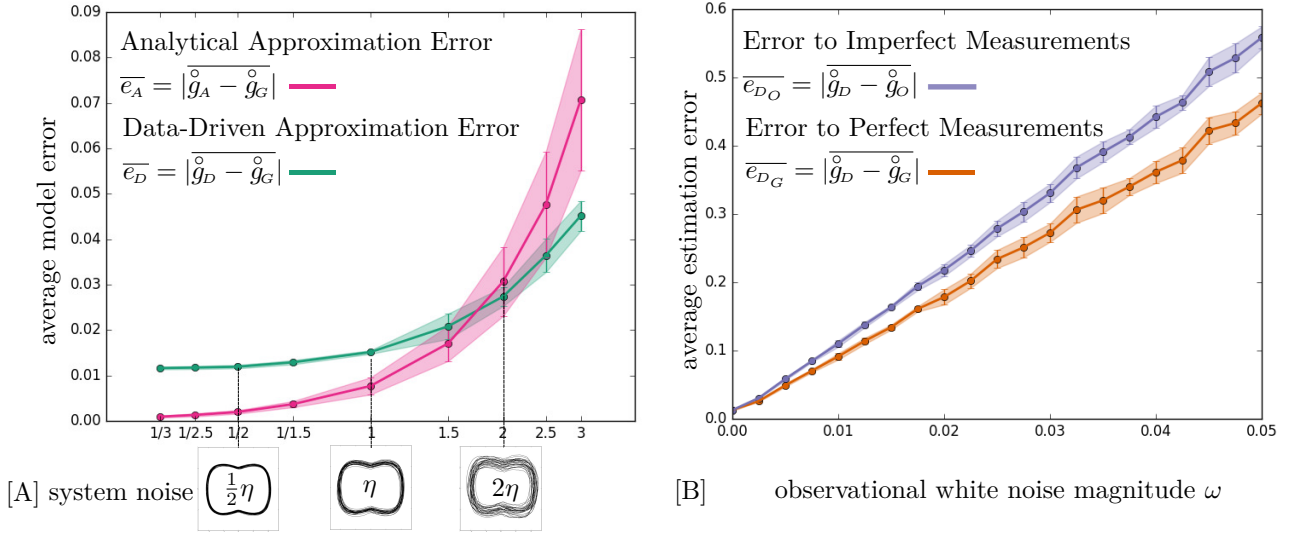}\label{fig:obserr}
\caption{[A] Comparing model error for the data-driven and analytical approximations. %
In simulation, the nine link swimmer is attracted to the same input gait in Fig. \ref{fig:gait} while experiencing a variety of system noise regimes (example trajectories are shown for $0.5\eta$, $\eta$, and $2\eta$). %
The accuracy of both models (data-driven in green, analytical in red) are plotted to show the estimation error in $x$ velocity over 20 trials. %
System noise degrades the model accuracy of both approaches. %
The data driven model retains accuracy at high noise levels at the expense of having poor accuracy in low noise regimes. %
[B] Comparing model error across a variety of observational error regimes. %
 In simulation, the nine link swimmer experiences the same input gait and attraction laws of Fig. \ref{fig:gait}. %
 We plotted the range of estimation error observed for the data-driven model under observation error $\omega$ with respect to (the ground truth model in orange and the observation data in purple) for the $x$ velocity over an ensemble of 10 trials at each observation error regime. %
 Although the data-driven connection is estimated with respect to the data indicated by the purple error metric, it shows a closer likeness to the ground-truth dynamics indicated by the orange error metric. %
 This suggests that the geometric constraints of the regression help mitigate some of the accuracy degradation that occurs during magnification of the observation noise.}
\end{figure}

The data-driven approach yields better models than the analytic Taylor expansion of the dynamics around the gait cycle when the system noise is more aggressive.
The effect is illustrated in Fig.~\ref{fig:obserr}, which shows estimation error for both methods as function of noise regime about the extremal gait.
The data-driven model also yields a better model in regions where the connection is highly nonlinear, as for the nine-link swimmer at the right of Fig.~\ref{fig:gait}(C).

These differences stem from the fact that the analytic model is computed from linearizations at specific points on the cycle, and the data-driven model is computed by minimizing approximation error across the neighborhood of the cycle.
At the limit of large samples and small noise, the data-driven model approaches the accuracy of the analytic model.
Thus, at the limit for many samples and finite noise, the data-driven model should always yield a better model than the analytic model. It should always give the best linearization for prediction over the available data, an improvement from the linearization locally at the gait cycle.
However, with finite sample sizes the accuracy of analytical model can out-perform the data-driven model.

Now white gaussian noise is added to the observation model, which simulates measurement error in a motion capture system. Fig. ~\ref{fig:obserr} shows that the modeling error does grow with the magnitude of the gaussian observation error.
However, the modeling error with respect to ground truth grows at a slower rate than what was observed.
This indicates that the system is able to reason about some of the dynamical structure that is occluded by the noise.

\section{Conclusions, Limitations, and Future Work}

The main contribution of this work is a method to compute local connections and cost metrics in the neighborhood of gait cycles, based solely on the observation of noisy trajectories.

The data-driven modeling approach relied strongly on the intrinsic noise of the system to produce sufficient excitations.
These excitations allowed for construction of a regression and identification of the structure of the dynamics at every phase of the cycle.
The strength of this approach is that the noise in magnitudes that have been observed in animals and robots can be exploited to model behaviors.
The weakness is a reliance on the noise being strictly dynamical rather than observational.
Measurement noise could mask some of the structure we expose by regression, as was shown in Fig. \ref{fig:obserr}.

The empirical savings for computation of local models over global models grow exponentially with shape dimension of the system observed.
This lays a basis on which to study motion modeling on complex platforms with far lower logistical overhead.

A goal of future work is to extend this method to a broader class of data-driven models outside those systems that have a connection-like structure \citep{ostrowski1998geometric_generalizedmomentum, bazzi2017motion}.
Improvements to the phase estimator, state estimator, or regression could strengthen the results presented here.

A promising extension of this work is to use these local models as tools to inference about gait improvements for biological and robotic systems.

\thanks{RLH thanks the National Science Foundation for support under CMMI grant 1653220. SR and BB were funded by ARO grant W911NF-14-1-0573 and the Rackham Merit Fellowship.}


\end{document}